\documentclass[11pt,twoside]{article}
\usepackage{./asp2014}
\usepackage{natbib}
\newcommand\cir{BS\,Cir}
\newcommand\vir{CU\,Vir}
\newcommand\ori{V901\,Ori}
\newcommand\sig{$\sigma$\,Ori\,E}

\resetcounters

\begin{document}

\title{On the nature of rotational period variability of magnetic stars}
\author{Zden\v{e}k Mikul\'a\v{s}ek$^{1}$, Ji\v{r}\'{i} Krti\v{c}ka$^1$, Jan Jan\'{i}k$^1$, Gregory W. Henry$^2$, Miloslav Zejda$^1$, Matthew Shultz$^3$, Ernst Paunzen$^1$, and Miroslav Jagelka$^1$}
\affil{$^1$Department of Theoretical Physics and Astrophysics,
Masaryk University, Kotl\'a\v{r}sk\'a 2, 611 37 Brno, Czech Republic; \email{mikulas@physics.muni.cz}}
\affil{$^2$Center of Excellence in Information Systems, Tennessee
State University, Nashville, Tennessee, USA}
\affil{$^3$Department of Physics and Astronomy, Uppsala University, Box 516, Uppsala 75120, Sweden}

\begin{abstract}
The magnetic chemically peculiar (mCP) stars of the upper main sequence exhibit periodic light, magnetic, radio, and spectroscopic variations that can be adequately explained by a model of a rigidly rotating magnetized star with persistent surface structures. The majority of mCP stars rotate at strictly constant periods. However, there are a few mCP stars whose rotation periods vary on timescales of decades while the shape of their phase curves remains unchanged. In the case of CU Vir and V901 Ori, we have detected {it\ cyclic} period variations. We demonstrate that the period oscillations of CU Vir may be a consequence of the interaction of the internal magnetic field and differential rotation.
\end{abstract}

\section{Introduction}

Magnetic chemically peculiar (mCP) stars are the most suitable test beds for studying rotation and its variation in upper (B2V to F6V) main-sequence stars. The surface chemical composition of mCPs is very uneven. Overabundant elements are, as a rule, concentrated into large spots persisting for decades to centuries. The abundance inhomogeneities on the stellar surface influence the star's spectral energy distribution.  As the star rotates, periodic variations in the brightness, spectrum, and magnetic field are observed.  We have studied both present and archival observations of all kinds to check the stability of mCP star rotation periods with impressive accuracy.

The changes in the periods were derived from shifts of (light, spectroscopic) phase curves obtained in the past. \citet{mik901} developed this method and applied it to helium strong \ori. Then it was many times improved and tested on mCPs and other types of variables \citep[see e.g.][]{mikecl,mikzej}. The method is based on the usage of suitable phenomenological models of phase curves and the period variation.  Solution through robust regression provides us with all model parameters and estimations of their uncertainty.

The vast majority of CP stars studied to date display strictly periodic variations. However, a few mCP stars, including \vir\ and \ori, have been discovered to exhibit rotational period variations.

\section{Rotationally variable magnetic stars}

\subsection{Famous silicon star CU Virginis}

CU Vir = HD 124224, is a bright, rapidly-rotating ($\overline{P}=0.520694$\,d), medium-aged silicon mCP star \citep{kochba}. It is also the first known hot stellar radio pulsar \citep[][and references therein]{trig}. \citet{pyp}, using their new and archival photometry, constructed an O-C diagram showing a sudden period increase of 2.6 s (slower rotation) in 1984! Another smaller jump toward a longer period in 1998 was reported by \citet{pypad}.  \citet{mikCU} processed all available measurements of \vir\ and found that its rotation was gradually slowing until 2005 and since then has been accelerating.

\begin{figure}
\centering\includegraphics[width=0.95\textwidth]{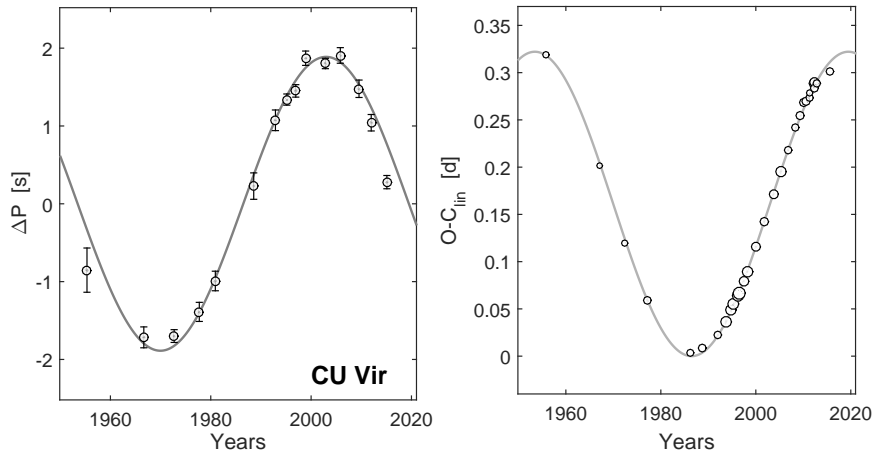}
\caption{(a) Changes in the rotation period of \vir\ in seconds with the respect to the mean rotation period $\overline{P}=0.520694$\,d. The period variations are approximated by a sinusoid reaching its extrema in 1970 and 2003. The amplitude of period variations is 3.78\,s. (b) O-C$_{\rm{lin}}$ variations versus the linear ephemeris at the time $T_0=1986.49(4)$ have semiamplitude $0.1611(5)$\,d\,$=0.31\,\overline{P}$.}\label{duoCU}
\end{figure}

We have collected 18617 new and archival measurements of \vir\ covering 67 years, including photometry in the wavelength range 200-753 nm, spectroscopy, radiometry, and magnetic field measurements. The period variations appear to be cyclic and may be periodic (see Fig.\,\ref{duoCU}). We derive a cycle timescale of $66.0\pm0.4$\,yr. \citet{krt} show that such period oscillations might result from the interaction of the internal magnetic field and differential rotation and predict a rotational cycle timescale of 51\,yr.

\subsection{Magnetic helium strong star V901 Orionis}

\ori\,= HD 37776 is a very young hot star (B2\,IV) with a complex global magnetic field \citep[][and references therein]{koc}. It is listed among He-strong CP stars, although its light variations are caused by the incidence of spectroscopic spots with an overabundance of silicon \citep{krt901}. \citet{mikjap} reported possible gradual changes in the stellar rotation. This suspicion was confirmed by \citet{mik901,mikCU} who found that the rotational deceleration was recently replaced by acceleration.

\begin{figure}
\centering\includegraphics[width=0.95\textwidth]{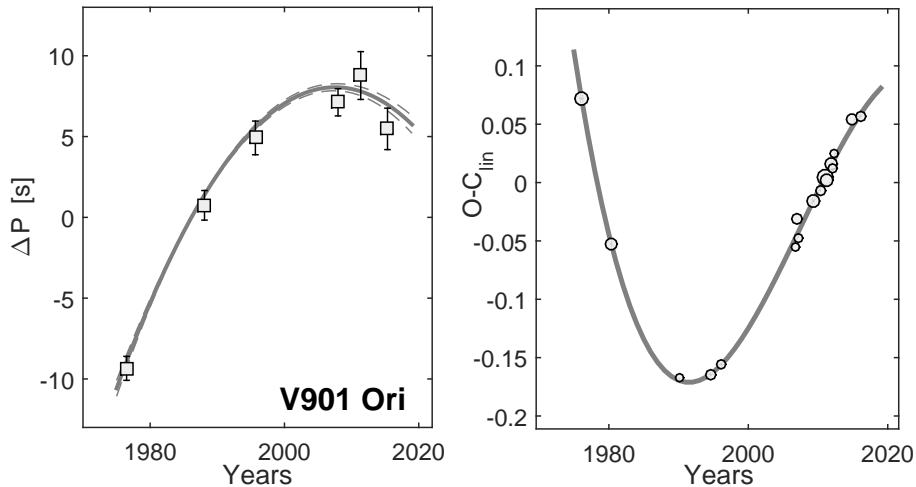}
\caption{(a) Changes in the rotational period of \ori\ in seconds. Time dependence of the period is approximated by a parabola reaching its maximum in 2009. (b) Changes in times of the zero phase in days can be well fitted by a cubic parabola.}\label{duo901}
\end{figure}

Analysing all available data, we conclude that the maximum period $P(t)$ occurred in 2007.7(1.3), when the second derivative of the period $\ddot{P}=-3.05(22) \times 10^{-12} $\,d$^{-1}$. If the observed O-C variation is a segment of cyclic changes, then the duration of the cycle should be longer than 120 years. Currently, we are not able to explain this type of the rotational variability; even the new mechanism described in \citet{krt} predicts an unacceptably short cycle for V901~Ori.

\subsection{Be/Helium strong star $\sigma$\,Orionis E}

$\sigma$\,Ori\,E = HD 37479 = V1030 Ori is a hybrid of a classical He-strong mCP star and a Be star with strong stellar winds. It is an extremely young and massive star with a period of photometric and spectral changes $P=1.19801$\,d. \citet{town} revealed the period to be increasing at the rate $\dot{P}=2.4(1)\times 10^{-9}=0.076$\,s\,yr$^{-1}$. The spin-down time \citep{unstead} $\tau=P/\dot{P}=1.4\times10^6$\,years is compatible with the age of the star. Consequently, the observed lengthening of the period can be explained by magnetic braking through strong stellar winds \citep{town}.

\subsection{Cool chemically peculiar star BS Circini}

\cir\,= HD 125630 is a relatively old \citep[$510^{\ \,+90}_{-150}$\,Myr,][]{kochba} SrCrEu mCP star with strong antiphase light variations in $y$ and $v$ colors \citep{mikzoo,mikBS}. \citet{mikmos,mikBS} revealed that the rotational period of $2.2042849(4)$\,d is increasing at the rate $\dot{P}=5.4(4) \times 10^{-9} = 0.181(13)$ s\,yr$^{-1}$. The spin-down time $\tau=P/\dot{P}=1.05(8)\times10^6$\,years represents only $0.2\,\%$ of the estimated age of the star. Total variations in O-C are smaller than $0.11 P$, so they might be caused by stellar precession \citep[][and references therein]{mik901,mikBS}.

\section{Conclusions}

We conclude that the rotational periods of a few mCP stars are gradually changing. However, the stars exhibiting these variations differ in almost all relevant aspects. It is very likely the period variations have more than one cause.

Single star evolution with standard stellar winds results in rotation changes too slow to detect. The light-time effect leads to radial velocity variations, but these are not detected.  Magnetic braking through the angular momentum loss caused by the wind escaping from the extended magnetosphere can apparently be effective only in very specific cases of hot mCPs as \sig. The precession of a magnetically distorted star might be taken into our account for \cir\, which has small-amplitude period changes. Period oscillations in \vir\ can be interpreted as a consequence of torsional waves that may disseminate within this magnetic rotating star \citep{krt}. Unfortunately, a plausible explanation for the rotational period variations in \ori\ remains elusive.

\acknowledgements The study was supported by the grant GA\v{C}R 16-01116S. GWH acknowledges support from NASA, NSF, and the State of Tennessee through its Centers of Excellence program.


\begin{thebibliography}  
\bibitem[Kochukhov \& Bagnulo(2006)]{kochba} Kochukhov, O. \& Bagnulo, S. 2006, \apj, 726, 24
\bibitem[Kochukhov et al.(2011)]{koc} Kochukhov, O., Lundin, A., Romanyuk, I., Kudryavtsev, D. 2011, \apj, 726, 24
\bibitem[Krti\v{c}ka et al.(2016)]{krt} Krti\v{c}ka, J., Mikul\'a\v{s}ek, Z., Henry, G. W., Kurf\"{u}rst, P., \& Karlick\'y, M. 2016, \mnras, in press
\bibitem[Krti\v{c}ka et al.(2007)]{krt901} Krti\v{c}ka, J., Mikul\'a\v{s}ek, Z., Zverko, J., \& \v{Z}i\v{z}\v{n}ovsk\'y, J. 2007, \aap, 470, 1089
\bibitem[Mikul\'a\v{s}ek(2015)]{mikecl} Mikul\'a\v{s}ek, Z. 2015,
    \aap, 584, A8
\bibitem[Mikul\'a\v{s}ek et al.(2015)]{mikBS} Mikul\'a\v{s}ek, Z., Jan\'ik, J., Krti\v{c}ka, J., Zejda, M., \& Jagelka, M. 2015, in ASP Conf. Ser. 494, Physics and Evolution of Magnetic and Related Stars, ed. Y. Y. Balega, I. I. Romanyuk, \& D. O. Kudryavtsev, 189
\bibitem[Mikul\'a\v{s}ek et al.(2008)]{mik901} Mikul\'a\v{s}ek, Z.,
    Krti\v{c}ka, J., Henry, G. W., Zverko, J., \v{Z}i\v{z}\v{n}ovsk\'y, J. et al.
    2008, \aap, 485, 585
\bibitem[Mikul\'a\v{s}ek et al.(2011a)]{mikCU} Mikul\'a\v{s}ek, Z., Krti\v{c}ka, J., Henry, G. W. et al. 2011a, \aap, 534, L5
\bibitem[Mikul\'a\v{s}ek et al.(2011b)]{unstead} Mikul\'a\v{s}ek, Z., Krti\v cka, J., Jan\'ik J. et al. 2011b, in Magnetic Stars, Proceedings of the International Conference, SAO RAS 2010, eds: I. I. Romanyuk and D. O. Kudryavtsev, 52
\bibitem[Mikul\'a\v{s}ek et al.(2014)]{mikmos} Mikul\'a\v{s}ek, Z., Krti\v{c}ka, J., Jan\'ik, J., Zejda, M., Henry, G. W., Paunzen, E., \v{Z}i\v{z}\v{n}ovsk\'y J., \& Zverko, J. 2014, in Putting A Stars into Context: Evolution, Environment, and Related Stars, ed. G. Mathys, E. Griffin, O. Kochukhov, R. Monier, G. Wahlgren, 270
\bibitem[Mikul\'a\v{s}ek et al.(2007a)]{mikjap} Mikul\'a\v{s}ek, Z.,
    Krti\v{c}ka, J., Zverko, J., \v{Z}i\v{z}\v{n}ovsk\'y, J., \& Jan\'ik, J. 2007, in ASP Conf. Ser. 361, Active OB-Stars: Laboratories for Stellar and Circumstellar Physics, ed. S. \v{S}tefl, S. P. Owocki, \& A. T. Okazaki, 466
\bibitem[Mikul\'a\v{s}ek \& Zejda, 2013]{mikzej} Mikul\'a\v{s}ek, Z. \&
    Zejda, M., in \'Uvod do studia prom\v{e}nn\'{y}ch hv\v{e}zd,
    ISBN 978-80-210-6241-2,  Masaryk University, Brno 2013
\bibitem[Mikul\'a\v{s}ek et al.(2007b)]{mikzoo} Mikul\'a\v{s}ek, Z., Zverko, J., Krti\v cka, J. et al. 2007, in Physics of Magnetic Stars, Proceedings of the International Conference, SAO RAS 2010, eds: I. I. Romanyuk and D. O. Kudryavtsev, Special Astrophysical Observatory, 300
\bibitem[Pyper \& Adelman(2004)]{pypad} Pyper, D. M., \& Adelman, S. J. 2004,
    The A-Star Puzzle, IAU Symposium No. 224, eds. J. Zverko, J.
    \v{Z}i\v{z}\v{n}ovsk\'{y}, S.\,J. Adelman, \& W.\,W. Weiss (Cambridge
    University Press, Cambridge), 307
\bibitem[Pyper et al.(1998)]{pyp} Pyper, D. M., Ryabchikova, T., Malanushenko, V. et al. 1998, \aap, 339, 822
\bibitem[Townsend et al.(2010)]{town} Townsend, R. H. D., Oksala, M. E.,
    Cohen, D. H., Owocki, S. P., ud--Doula, A., 2010, \apj, 714, 318
\bibitem[Trigilio et al.(2000)]{trig} Trigilio, C., Leto, P., Leone, F., Umana, G., \& Buemi, C. 2000, \aap, 362, 281




\end{thebibliography}
\end{document}